\documentclass[11pt]{article}
\usepackage{amsmath}
\usepackage{amssymb}
\usepackage{graphicx}
\usepackage[margin=2cm]{geometry}
\usepackage{dcolumn}
\usepackage{bm}
\usepackage{subfigure}
\usepackage{color}
\usepackage[font=small]{caption}

\newcommand{\FeCo}[2]{Fe$_{\mathrm{#1}}$Co$_{\mathrm{#2}}$}

\begin{document}

\begin{center}
{\LARGE\bfseries Temperature and crystallographic orientation dependence of the anisotropic magnetoresistance in epitaxial Fe$_{65}$Co$_{35}$ thin films}
\end{center}
\vspace{0.1cm}
\begin{center}
A. Paz Jalca$^{1,+}$, W. H. Painado Lozano$^{1,+}$, D. E. Gonzalez-Chavez$^{2}$,  L. Saba$^3$, D. Pérez-Morelo$^{3,4}$, J. E. Gómez$^{4,5}$, A. Butera$^{3,4,5}$, A. Gutarra Espinoza$^{1}$,  L. M. Leon Hilario$^{1}$, L. Avilés-Félix$^{3,4,5,*}$
\vspace{0.5cm}

$^1$Facultad de Ciencias, Universidad Nacional de Ingeniería, Lima, Peru\\
$^2$Departamento de Ciencias Naturales, Universidad Católica San Pablo, Arequipa 04001 Peru.\\
$^3$Instituto Balseiro, Universidad Nacional de Cuyo, Comisión Nacional de Energía Atómica, R8402AGP Bariloche, Río Negro, Argentina\\
$^4$Instituto de Nanociencia y Nanotecnología (CNEA - CONICET), Nodo Bariloche,(8400) Rio Negro, Argentina\\
$^5$Departamento Magnetismo y Materiales Magnéticos, Gerencia Física, Comisión Nacional de Energía Atómica, R8402AGP Bariloche, Río Negro, Argentina\\
$^*$Corresponding author: luis.aviles@ib.edu.ar\\
$^+$These two authors contributed equally\\
\end{center}

\begin{abstract}
In this work, we study the anisotropic magnetoresistance (AMR) behavior of [001] epitaxial \FeCo{65}{35} thin films along different crystallographic directions as a function of temperature. The AMR ratio is found to strongly depend on the current orientation relative to the crystal axes, reaching 0.16\% and 0.10\%  at room temperature when the current is applied along the magnetic hard and easy axes, respectively. Moreover, the AMR ratio decreases at different rates as the temperature is reduced to 80 K. The longitudinal and transverse magnetoresistance curves were fitted using the Stoner–Wohlfarth formalism to describe the magnetization reversal path and to extract the magnetic anisotropy constants. The fitted cubic and uniaxial anisotropy constants are $K_c = -2.36$ kJ/m$^3$ and $K_u = 2.18$ kJ/m$^3$, verifying the change in the cubic anisotropy compared to Fe-richer \FeCo{100-x}{x} compositions. These results demonstrate that by tailoring the crystalline orientation and temperature dependence of AMR, epitaxial \FeCo{65}{35} thin films can enable the design of magnetic sensors with tunable sensitivity.
\end{abstract}

\section{\label{sec:Intro}Introduction}
Magnetotransport phenomena in ferromagnetic layers are of great significance for the design and development of magnetic sensors and spintronic devices. In particular, anisotropic magnetoresistance (AMR)—the dependence of the electrical resistance on the relative angle between the directions of the applied current and the magnetization—has played a major role in advancing modern magnetic technologies. Its applicability spans a wide range of fields, including magnetic field sensing \cite{LIN2025}, green microelectronics \cite{Wang2024}, wearable and flexible sensors \cite{Solis2025, NICOLICEA2025, Cheng2025, Bo2025}, and automotive electronics \cite{Miklusis2022, Balamutas2023}. Within this framework, \FeCo{100-x}{x} alloys have emerged as promising candidates for spintronic devices and magnetic sensors. Their tunable magnetic anisotropy and exceptionally low damping parameters, observed in both polycrystalline and epitaxial layers, are closely correlated with their magnetotransport and spin-transport properties \cite{Li2019, Velazquez2021, Saba2025}.

Among metallic ferromagnets, Fe-rich \FeCo{100-x}{x} alloys exhibits the lowest reported intrinsic damping values \cite{Mankovsky2010, Schoen2016, Natale2021}, with recent studies demonstrating that the damping coefficient depends on the crystallographic orientation \cite{Li2019, Zeng2020a, Velazquez2024}. This anisotropic damping arises from the variation of spin–orbit coupling (SOC) along different crystallographic directions of the cubic lattice. The SOC induces mixing between spin-up and spin-down $d$ states, which depends on the magnetization direction and consequently modulates the density of unoccupied $d$ states at the Fermi level. The variation of the SOC also affects the $s$–$d$ electron scattering cross-section and is responsible for the AMR in ferromagnetic metals \cite{Thomson1856}. Recent works on \FeCo{100-x}{x} alloyed thin films have further highlighted the role of anisotropic damping in epitaxial \FeCo{100-x}{x} layers \cite{Li2019, Zeng2020a, Velazquez2024}, and how it impacts both the spin-transport and magnetotransport properties of polycrystalline and epitaxial films \cite{Berger1988, Ganguly2014, Haidar2015, Weber2019, Zeng2020}. A detailed understanding of the AMR contributions due to the crystalline structure is therefore crucial for accurately quantifying the underlying magnetotransport parameters.

The angular dependence of the resistance in polycrystalline ferromagnetic materials can be derived from the generalized Ohm’s law \cite{1960Juretschke, Harder2016}. Considering a magnetic layer with in-plane magnetization, and assuming that AMR is the dominant contribution to the magnetoresistance, the induced electric field can be expressed as $\mathbf{E} = \rho_\perp \mathbf{j} + (\rho_\parallel - \rho_\perp)\mathbf{m}\left(\mathbf{m} \cdot \mathbf{j}\right)$, where $\mathbf{j}$ is the current density, and $\mathbf{m}$ is the unit vector in the direction of magnetization. From this expression we can define the AMR ratio as:
\begin{equation}
\mathrm{AMR}_\mathrm{ratio}=\frac{\Delta \rho}{\rho_\perp} = \frac{\rho_\parallel - \rho_\perp}{\rho_\perp}.
\label{Eq:AMR}
\end{equation}

On the other hand, in ferromagnetic single crystals or epitaxial films, electron scattering is strongly influenced by the orientation of the current relative to the crystallographic axes, resulting in an angular dependence of the anisotropic magnetoresistance. Because both the magnetic damping and the anisotropic magnetoresistance originate from spin-orbit coupling, a crystalline-dependent AMR is naturally expected. Experimental evidence of such crystalline-dependent AMR, arising from anisotropy in the spin–orbit coupling, has been recently reported by different authors in several \FeCo{}{} alloys \cite{Li2019, Zeng2020a,Zeng2020,ZENG2020c}. 

In this work, we investigate the influence of crystal symmetry on the magnetotransport properties of epitaxial \FeCo{65}{35} films, with particular emphasis on the anisotropic magnetoresistance and its dependence on the current, external magnetic field orientations and the temperature. The epitaxial growth of the \FeCo{65}{35} layers on MgO substrates ensures well-defined crystalline order, allowing us to probe the interplay between crystal symmetry and magnetotransport behavior. We explore the correlation between the magnetization process, the magnetotransport and the underlying crystallographic structure by systematically measuring the resistance while varying the directions of current and magnetization with respect to different crystal axes. By analyzing the angular dependence of the magnetoresistance and fitting the experimental data using the generalized Ohm’s law and free energy minimization, we extract the uniaxial and cubic anisotropy constants and use them to compute the longitudinal and transverse voltage responses for different current and field orientations. Finally, we investigated the temperature dependence of the resistivity and the AMR ratio for currents applied along two different crystallographic directions. The effect of temperature on the magnetoresistive response provides valuable insight into the thermal stability and thermo-sensitivity of the system, both of which are critical parameters for potential sensor applications \cite{Tamulynas2026, Ripka2013, Lim2022}.

\section{\label{sec:Experimental} Experimental Details}
\begin{figure}[ht]
\centering
\includegraphics[width=0.75\linewidth]{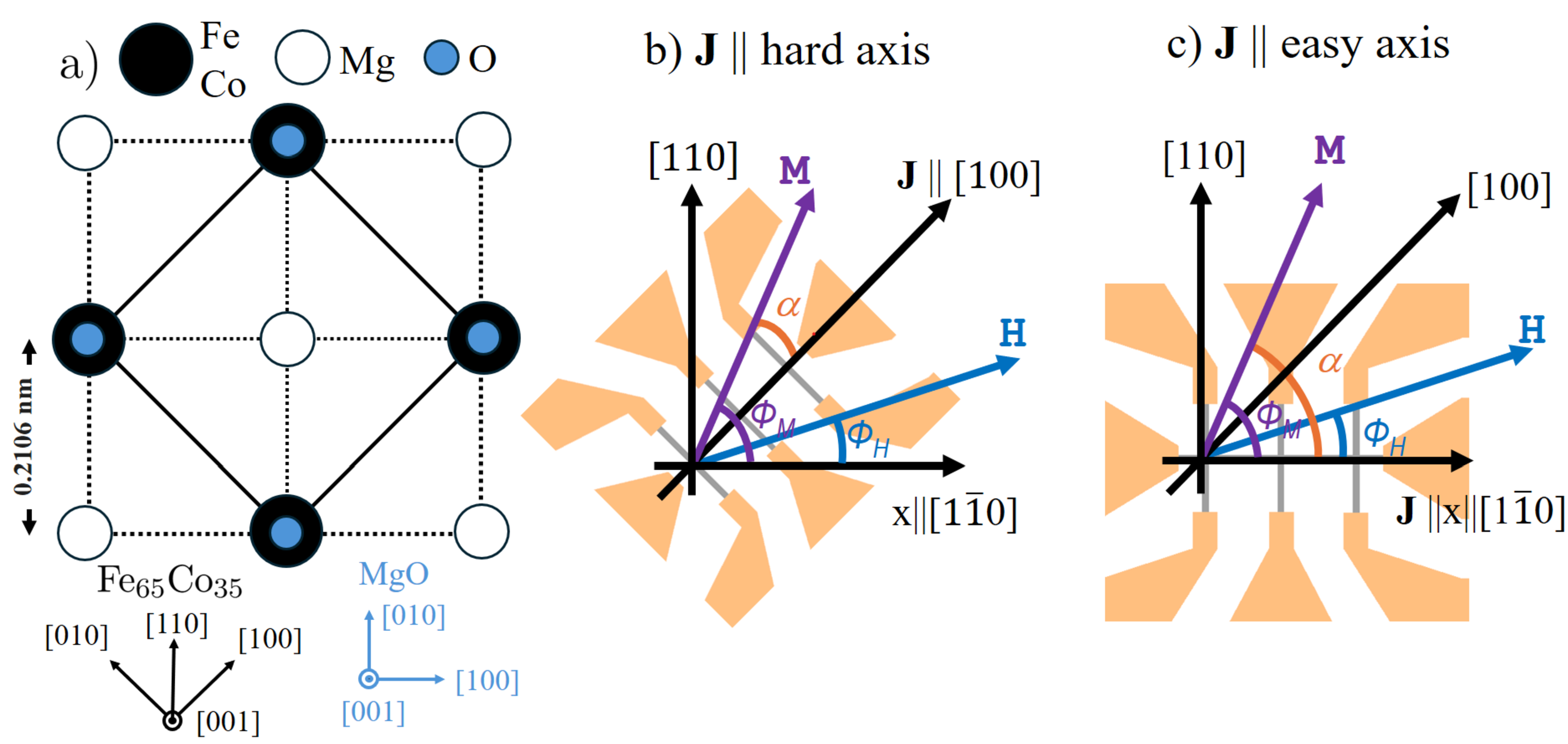}
\caption{a) Schematic representation of the atomic arrangement of \FeCo{65}{35} grown epitaxially on MgO. The crystallographic directions of the MgO and of the \FeCo{65}{35} are indicated at the bottom in light blue and in black, respectively. The [100] direction of the \FeCo{65}{35}, which corresponds to a hard axis, is rotated by 45$^\circ$ with respect to the [100] of the MgO. Representation of the angles of the magnetization ($\phi_M$) and magnetic field vector ($\phi_H$) measured from the $x$-axis and of the Hall bar setups where the current flows along the b) hard axis [100] and c) easy axis [1$\overline{1}$0], respectively.} 
\label{fig:scheme}
\end{figure}

An \FeCo{65}{35}/Pt bilayer was deposited by magnetron sputtering on MgO (001) single crystal substrates. The \FeCo{65}{35} layer was deposited from an alloyed target with a substrate-to-target distance of 7.5 cm. The base pressure before depositing the samples was $\lesssim 0.13 \times 10^{-3}$ Pa and the growth was performed under an Ar pressure of 0.24 Pa, at a fixed power density of 1.32 W/cm$^{2}$. To improve the epitaxial growth of the \FeCo{65}{35}, the substrate was kept at a fixed temperature of 423 K during deposition. The protective Pt layer, on the other hand, was deposited at room temperature, under an Ar pressure of 0.35 Pa and a power density of 1.76 W/cm$^{2}$. The nominal layer thickness was fixed to 15 nm for \FeCo{65}{35} and 10 nm for Pt. The structural and compositional characterization of the samples grown under this parameters---performed using X-ray diffraction and X-ray reflectivity, scanning transmission electron microscopy with high angular annular dark field (STEM-HAADF), and electron energy-loss spectroscopy (EELS)---can be found in our previous works \cite{Velazquez2021, Velazquez2024,Velazquez2020}. Magnetization loops were measured using a magneto-optical Kerr effect magnetometer in the longitudinal configuration. The magnetic field was applied in the plane of the sample along the selected crystallographic directions. A 632 nm, 5 mW red laser beam was focused onto the sample surface, and the polarization of the reflected beam was analysed in order to monitor the component of the magnetization parallel to the applied field. The magnetotransport measurements were performed in a setup consisting of a Keithley 6221 current source and two HP34401A multimeters. Samples were mounted inside a Janis SVT-300 cryostat to control the temperature of the sample. The external magnetic field was applied using a Danfysik System 8000 power supply connected to a GMW electromagnet mounted on a goniometer that can rotate 360 degrees. The sample holder was oriented such that the magnetic field could rotate within the plane of the film. The axes corresponding to the crystalline directions of the \FeCo{65}{35} layer are schematically shown in Fig. \ref{fig:scheme}a.
The longitudinal and transverse voltages were measured simultaneously in Fe$_{65}$Co$_{35}$/Pt micro-sized Hall bars made using standard optical lithographic techniques. The details of the Hall bars, such as the separation between the contacts and dimensions of the bar, can be found in the Supplementary Data file. Two Hall bars were used for the characterization of the magnetotransport properties, one oriented along the crystalline direction [100] of the \FeCo{65}{35} that correspond to the magnetization hard axis and the other bar oriented along the magnetization easy axis ([1$\overline{1}$0]). These two configurations are schematized in Figs. \ref{fig:scheme}b and \ref{fig:scheme}c. Our $x$-axis is fixed along the crystalline direction [1$\overline{1}$0] of \FeCo{65}{35}. The equilibrium magnetization angles $\phi_M$ and the magnetic field angle $\phi_H$ are always measured from the $x$-axis. $\alpha$, the angle between the current and the magnetization vector is also indicated in Figs. \ref{fig:scheme}b and \ref{fig:scheme}c. To determine the anisotropic magnetoresistance ratio, longitudinal $V_{xx}$ and transverse $V_{xy}$ voltages are measured by varying the field direction ($\phi_H$), while fixing the sample's temperature and the external field amplitude at 150 mT. Finally, measurements of $V_{xx}$ and $V_{xy}$ were performed by varying the magnetic field amplitude at fixed field directions $\phi_H$ of the external magnetic field and at different temperature values ranging from 80 K to 300 K.
 
\section{\label{sec:Results}Results}
The ferromagnetic alloy \FeCo{65}{35} grown on single-crystal MgO(001) substrates exhibit cubic crystal symmetry with a bcc structure \cite{Velazquez2021,Velazquez2024,Velazquez2020}. By keeping the temperature of the MgO substrate constant at 423 K during the growth of the \FeCo{65}{35} we achieved layers with the [100] axis rotated by 45$^\circ$ with respect to the MgO [100] substrate direction, as schematically shown in Fig. \ref{fig:scheme}. In our previous work, we reported the structural characterization including high resolution transmission electron microscopy and $\phi-$scan X-ray diffraction measurements which verify the epitaxial growth of the FeCo \cite{Velazquez2024}. We also demonstrated that the \FeCo{65}{35} presents a magnetic hard axis, oriented toward the crystallographic [100] direction, in contrast to Fe-richer samples where the magnetic hard axis is oriented toward the crystallographic [1$\overline{1}$0] direction, which suggest a sign change in the cubic magnetocrystalline anisotropy constant $K_c$. This is schematically represented on Fig. \ref{fig:scheme} where we can identify that the cubic easy axes in our samples correspond to the directions parallel to [110] and [1$\overline{1}$0] and the hard axes are along the \FeCo{65}{35} [010] and [100] directions.

\subsection{\label{sec:Kerr}Kerr magnetometry}
\begin{figure}[ht]
\centering
\includegraphics[width=0.35\linewidth]{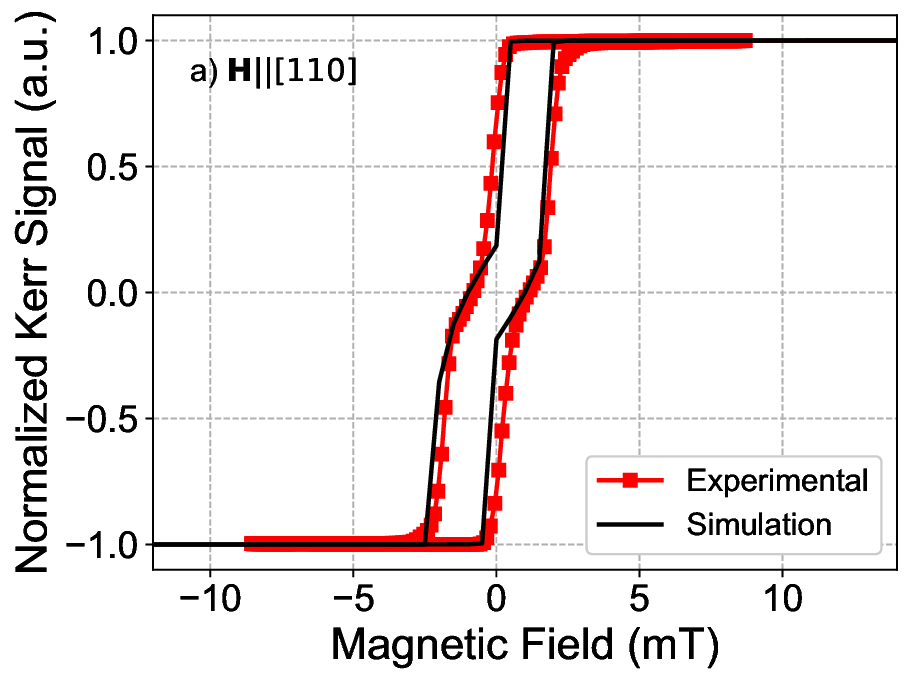}
\includegraphics[width=0.35\linewidth]{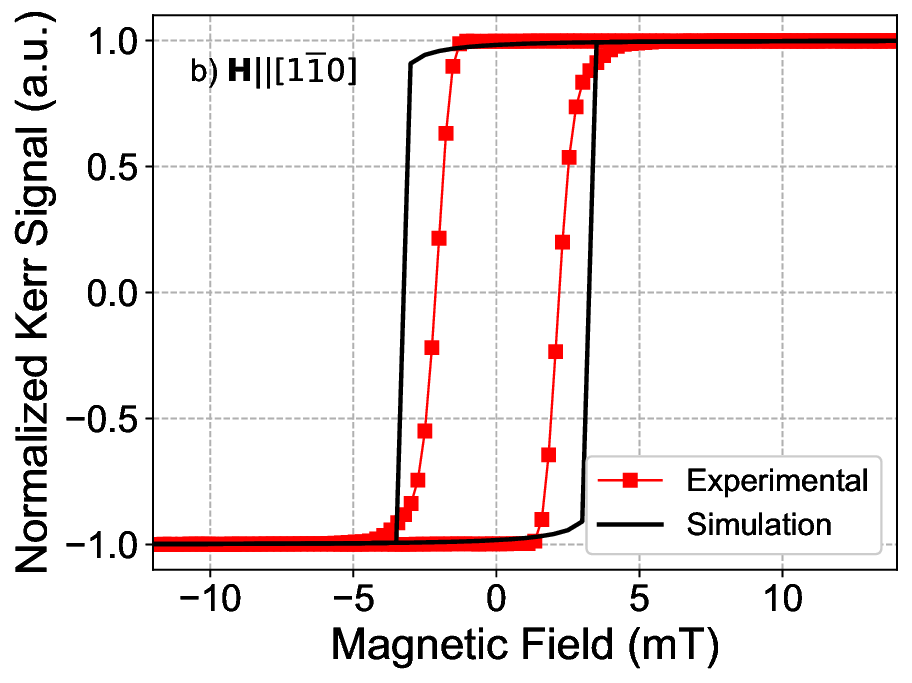}
\caption{Kerr hysteresis loop of the \FeCo{65}{35}/Pt bilayer with the external magnetic field applied along the a) [110] axis and b) [1$\overline{1}$0] axis. The solid lines correspond to the simulated hysteresis loops and the red squares indicate the experimental data.}
\label{fig:Kerr}
\end{figure}

The average value of the saturation magnetization of \FeCo{65}{35} layers is $M_s$ = 1600 kA/m \cite{Velazquez2024}. Kerr hysteresis loops with the external magnetic field applied along different crystallographic directions of the \FeCo{65}{35} layer ([1$\overline{1}$0], [010], [110], and [100]) confirm that the samples present a fourfold symmetry with the presence of a relatively weak uniaxial anisotropy. The saturation field is around 15 mT when the external magnetic field is applied close to the magnetization hard axes. Figure \ref{fig:Kerr} displays normalized Kerr loops measured with the external magnetic field applied along the [110] and [1$\overline{1}$0] directions of the \FeCo{65}{35} layer. Although both these directions correspond to the cubic easy axes of epitaxial \FeCo{65}{35}, the loops reveal the presence of an additional uniaxial anisotropy along  [1$\overline{1}$0], competing with the cubic contribution \cite{Saba2025,Velazquez2020}. 

The two-step magnetization reversal observed in Fig. \ref{fig:Kerr}a indicates that the magnetic field first overcomes the hard uniaxial anisotropy, whereas the second step—occurring near 2 mT—corresponds to the complete magnetization reversal once the cubic anisotropy is overcome. Figures \ref{fig:Kerr}a and \ref{fig:Kerr}b also present simulated hysteresis loops obtained by fitting the experimental data with a model based on the Stoner–Wohlfarth formalism that will be introduced in Section \ref{sec:Model} \cite{Saba2025}. Additional Kerr loops, measured with the external magnetic field applied along the [010] and [100] crystallographic directions, are provided in the Supplementary Data. The analysis of these loops and their corresponding fits enabled the extraction of the cubic and uniaxial anisotropy constants, which are subsequently employed to model the longitudinal and transverse resistance curves as a function of the external magnetic field angle. The observed dependence of the loop shape and coercive field on the crystallographic orientation is particularly significant for interpreting the magnetotransport measurements discussed in the following sections.

\subsection{\label{sec:Transport}Magnetotransport}
The magnetotransport experiments consist of measuring the longitudinal ($V_l$) and transverse ($V_t$) voltages with respect to the current direction. These voltages are then converted into resistivities using the geometrical dimensions of the Hall bar. The corresponding longitudinal and transverse resistivities can be modeled as:
\begin{eqnarray}
\rho_l = \rho_\perp + (\rho_\parallel - \rho_\perp)\cos^2\alpha, \label{Eq:rhoxx} \\
\rho_t = (\rho_\parallel - \rho_\perp)\cos\alpha \sin\alpha + \rho_\mathrm{offset}, \label{Eq:rhoxy}
\end{eqnarray}
where $\alpha$ is the angle between the magnetization and the current as shown in Figs. \ref{fig:scheme}b and \ref{fig:scheme}c. The offset term in the transverse resistivity is associated to a constant voltage contribution that does not originate from the AMR, but from experimental asymmetries of the Hall bar, such as geometrical misalignment between voltage probes, inhomogeneous current distribution, or a slight out-of-plane field component. This contribution is captured by the term $\rho_\mathrm{offset}$ in Eq.~\ref{Eq:rhoxy}. The dimensions of the bar used to convert resistance into resistivity were L$_l$ = 125 $\mu$m, L$_t$ = 10 $\mu$m, where L$_{l}$, L$_{t}$ are the length of the segment used for measuring the longitudinal $V_l$ and transversal $V_t$ voltages, respectively. The injected current in all the measurements was 1 mA which corresponds to a current density $J$ = 4 $\times$ 10$^{8}$ A/cm$^{2}$.

\subsubsection{Temperature dependence of anisotropic magnetoresistance ratio}

\begin{figure}[ht]
\centering
\includegraphics[width=\linewidth]{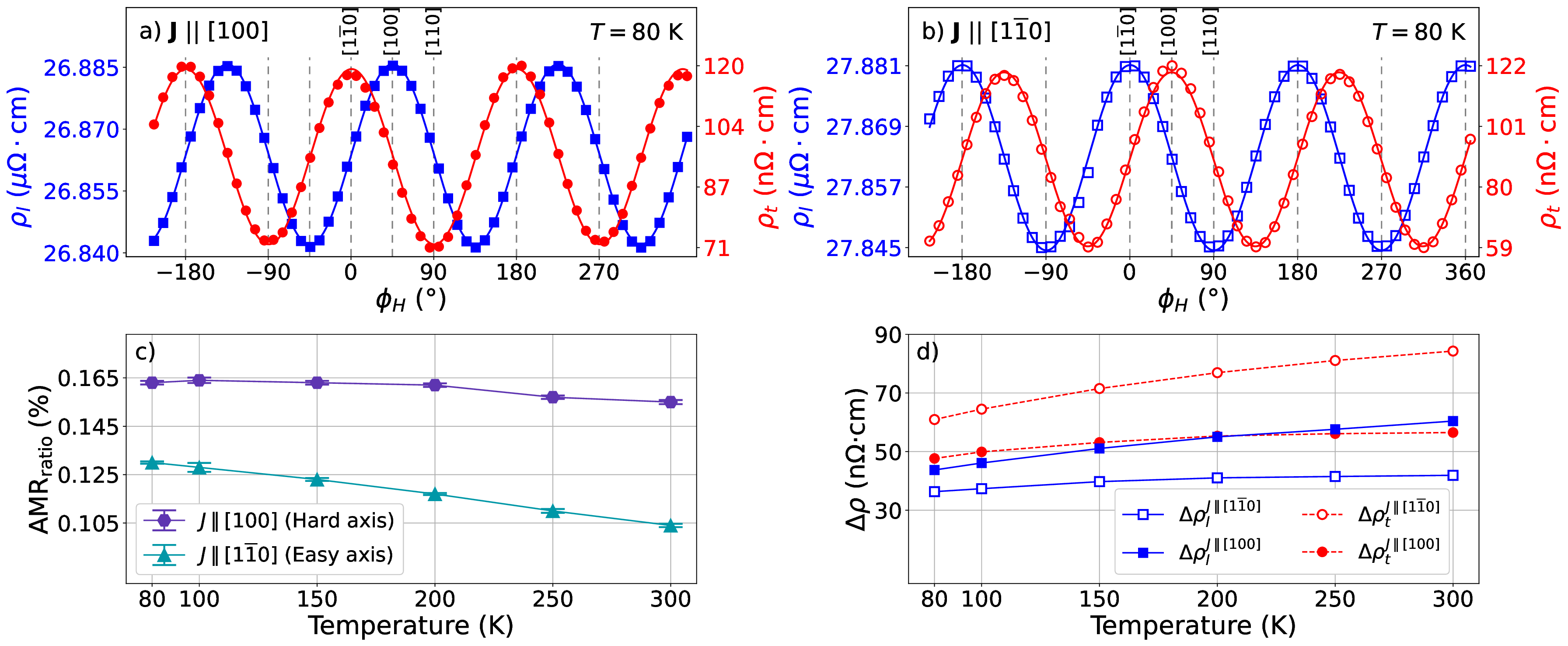}
\caption{Longitudinal $\rho_{l}$ (blue squares) and transverse $\rho_{t}$ (red circles) resistivity as a function of the external magnetic field direction ($\phi_H$) acquired at $T$ = 80 K. The current was applied along the \FeCo{65}{35} a) hard axis ([100]) and b) easy axis ([1$\overline{1}$0]). c) AMR ratio as a function of the temperature acquired with the current applied along the hard axis (purple circles) and easy axis (green triangles). d) $\Delta\rho = \rho_\parallel - \rho_\perp$ as a function of the temperature extracted from the angular dependence of $\rho_{l}$ and $\rho_{t}$ when the current is applied along [110] and [1$\overline{1}$0]. The current density applied was \textbf{J} = 4 $\times$10$^8$ mA/cm$^2 $}
\label{fig:AMRvT}
\end{figure}

Fig. \ref{fig:AMRvT}a shows the longitudinal $\rho_{l}$ and transverse $\rho_{t}$ resistances as functions of the magnetic field direction $\phi_H$, measured at 80 K with an applied external magnetic field of 150 mT, when the current is applied along the \FeCo{65}{35} [100] axis. $\rho_{l}^{[100]}$ reaches its maximum for \textbf{H}$\parallel$[100] ($\phi_H = 45^\circ$) and decreases as the external magnetic field is rotated toward directions transverse to the current. It is important to notice that $\phi_H$ is always measured from the [1$\overline{1}$0] direction as was indicated in Figs. \ref{fig:scheme}b and \ref{fig:scheme}c. On the other hand, the transverse resistivity curve ($\rho_{t}$) is shifted by 45$^\circ$, consistent with the $\sin\alpha\cos\alpha$ term in Eq.~\ref{Eq:rhoxy}. The AMR ratio was calculated from $\rho_\parallel$ and $\rho_\perp$ obtained by fitting Eq.~\ref{Eq:rhoxx} to the angular variations of $\rho_{l}$ measured at different temperatures. The corresponding resistivity plots are provided in the section S3 of the Supplementary Data File. 

For the current along the hard axis [100], the AMR ratios were 0.155 \% at $T = 300$ K and 0.163 \% at $T = 80$ K. The same analysis for the current along the easy axis [1$\overline{1}$0] yielded 0.104 \% and 0.130 \% at 300 K and 80 K, respectively. A crystal-axis-dependent anisotropic magnetoresistance has been also observed in our previous work for \FeCo{85}{15} \cite{Saba2025}. We reported in single epitaxial \FeCo{85}{15} thin films that the AMR ratio varies from 0.20 \%, when the current is applied along the \FeCo{85}{15} hard axis [110], to 0.17 \% when the current is applied along the easy axis [100], which represents a variation of the AMR ratio of 17 \%. In this work, the AMR ratio also decreases when we change the direction of the applied current from the hard axis (AMR$_\mathrm{ratio}^{[100]}$ = 0.155 \%) to the easy axis (AMR$_\mathrm{ratio}^{[1\overline{1}0]}$ = 0.104 \%) at room temperature, which represents a variation of 49 \%. A larger variation of the AMR in FeCo was expected in alloys with more Co content \cite{Saba2025,Zhang2025,Dong2023}. Our AMR measurements along [100] and [1$\overline{1}$0] confirm its dependence on the crystallographic direction of the current injection. This crystalline dependence arises from differences in the electronic structures along each direction \cite{Miao2024}. Similar behavior has been reported in several epitaxial \FeCo{}{} alloys, with AMR ratios varying in some cases up to $\approx$100 \% \cite{Saba2025,Zeng2020a,Haidar2015,Zeng2020,Tondra1993}.

The full temperature dependence of the AMR for both axes is shown in Fig.~\ref{fig:AMRvT}c. When the current is applied along the easy axis, the AMR increases almost linearly by about 30 \% with decreasing temperature, while along the hard axis it remains essentially constant. A linear temperature dependence of the AMR ratio has also been reported in ferromagnetic \FeCo{100-x}{x} alloys, showing the same behavior \cite{Natale2024}. In crystalline transition-metal \FeCo{}{} alloys, the AMR ratio decreases with increasing temperature due to a change in the dominant electron scattering mechanism, from impurities at low temperature to phonons at high temperature \cite{Berger1988,Freitas1990}. This behavior can also be interpreted within the formalism of Kokado \textit{et al}. \cite{Kokado2012}, where the resistivity from electron scattering by nonmagnetic impurities is attributed to spin mixing of localized $d$ states. In this framework, the spin-mixing resistivity term $\rho_{\uparrow\downarrow}(T)$ makes only a minor contribution, since it originates from electron–magnon or electron–electron scattering, which remains relatively weak at high temperatures \cite{Berger1988,1970Campbell}. It is also expected that the tetragonal distortions caused by the presence of Co atoms within the Fe cubic lattice play a relevant role in the magnitude of the spin–orbit coupling \cite{Li2019,Zhang2023, LEE2024} and, consequently in the anisotropy of the AMR ratio. One possible manifestation of this effect can be observed in the different values of $\Delta\rho$ obtained from the longitudinal and transverse resistivity drops, $\Delta\rho_l^{[1\overline{1}0]}$ and $\Delta\rho_t^{[1\overline{1}0]}$. A distorted unit cell induces a spin–orbit coupling strength that depends on the crystallographic direction, leading to distinct $\Delta\rho_l$ and $\Delta\rho_t$ values. Figure~\ref{fig:AMRvT}d shows the temperature dependence of $\Delta\rho_l$ and $\Delta\rho_t$. Here, a clear difference between both components is observed when the current is applied along the [1$\overline{1}$0] direction. In contrast, when the current is applied along the [100] direction, $\Delta\rho_l$ and $\Delta\rho_t$ show nearly identical values. This anisotropic behavior suggests that the spin–orbit coupling—and therefore the AMR amplitude—is sensitive to the underlying tetragonal distortion and the specific orientation of the current relative to the crystalline axes. It is also important to note that $\Delta\rho$ increases while the AMR ratio decreases as the temperature increases, indicating that the variation of the AMR ratio is primarily governed by the temperature dependence of $\rho(T)$.

\subsubsection{\label{sec:MR}Longitudinal and transverse resistivity measurements}

\begin{figure}[ht]
\centering
\includegraphics[width=0.39\linewidth]{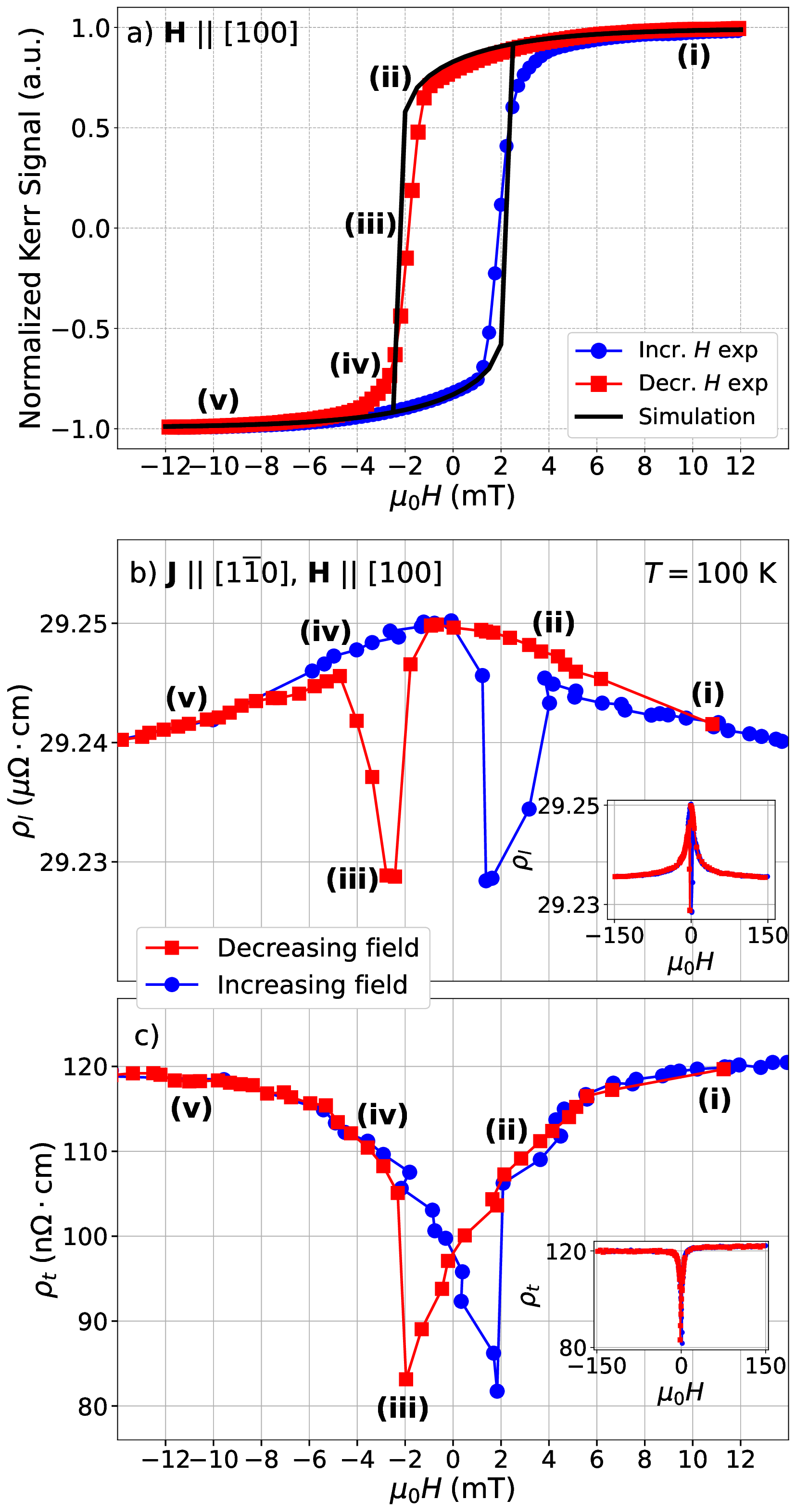}
\includegraphics[width=0.39\linewidth]{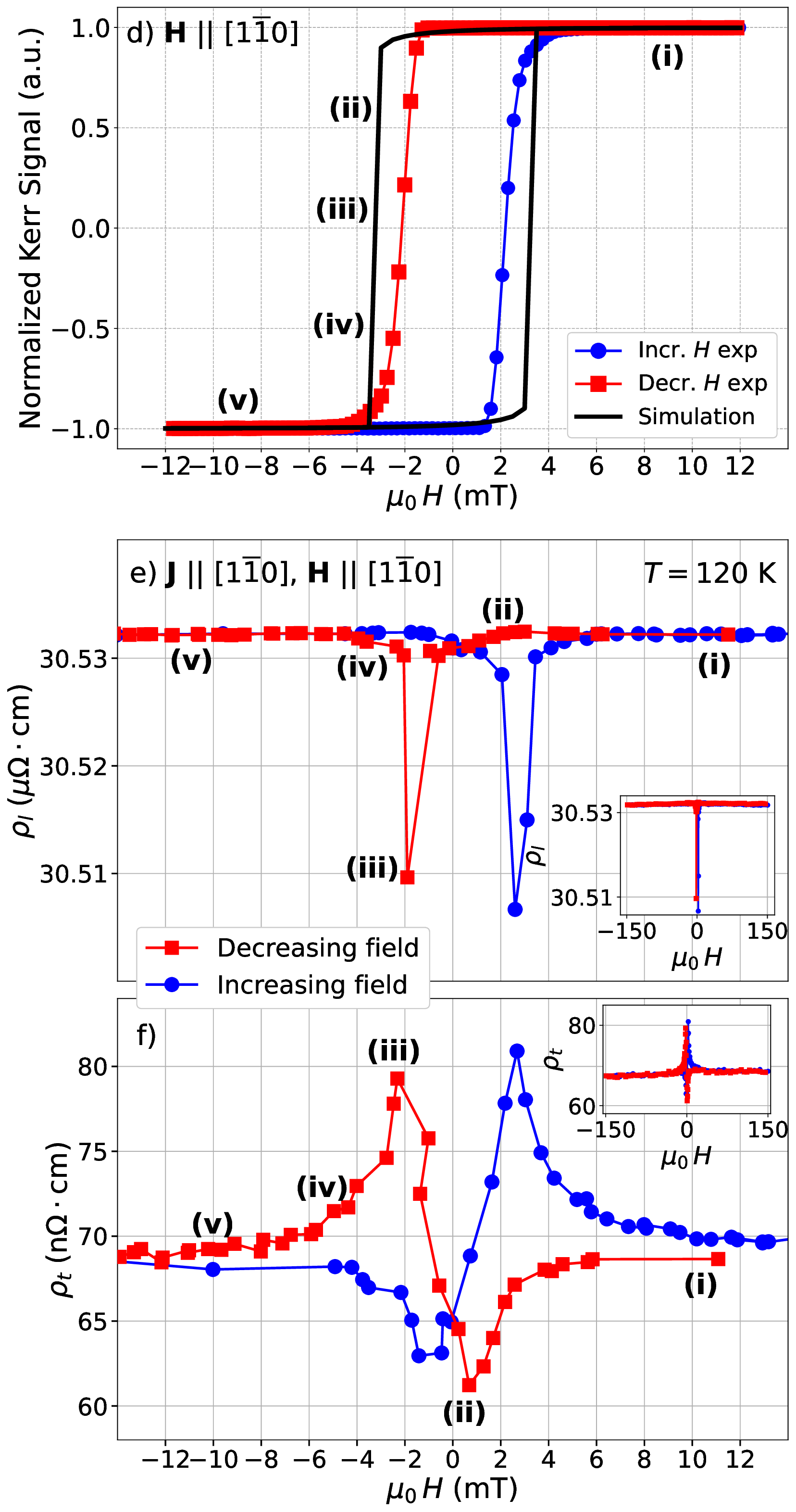}
\caption{Normalized hysteresis loop of the \FeCo{65}{35} layer acquired with the external magnetic field applied along the \FeCo{65}{35} a) [100] (hard) and d) [1$\overline{1}$0] (easy). The red squares and solid line indicate the data acquired when the external magnetic field is decreasing, and the blue circles and solid line correspond to the increasing of the external magnetic field. The simulated hysteresis loops are shown in a black solid line. b), e) $\rho_{l}$ and c), f) $\rho_{t}$ as a function of the external field when the current is applied along the [1$\overline{1}$0] axis and the external field is applied parallel to the hard axis [100] ($\phi_H =  45^\circ$) and along the easy axis ($\phi_H =  0^\circ$). The red squares and blue circles correspond to the data acquired when the external magnetic field is decreasing and increasing, respectively. The insets show the measurements of $\rho_{l}$($H$) and $\rho_{t}$($H$) in the range -150 mT to 150 mT. The data were acquired at 100 K and 120 K.}
\label{fig:RvH-J110}
\end{figure}

The magnetization process is further analyzed from the field dependence of resistivity. From Eqs. \ref{Eq:rhoxx} and \ref{Eq:rhoxy}, we can deduce that $\rho_{l} \propto M_{\parallel j}^2$ and $\rho_{t} \propto M_{\parallel j}M_{\perp j}$, with $M_{\parallel j}$ and $M_{\perp j}$ the magnetization components parallel and perpendicular to the current. Figure \ref{fig:RvH-J110} shows $\rho_l$ and $\rho_t$ measured with the current applied along the \FeCo{65}{35} easy axis [1$\overline{1}$0] and with the external magnetic field applied along the cubic hard axis [100] ($\phi_H = 45^\circ$) and the cubic easy axis [1$\overline{1}$0] ($\phi_H = 0^\circ$). For comparison, the normalized Kerr loop is also included to discuss the magnetization reversal process. By comparing the Kerr loop shown in Fig. \ref{fig:RvH-J110}a with those in Fig. \ref{fig:Kerr}, a characteristic loop shape is observed when the magnetic field is applied along the hard axis. The remanent Kerr signal in Fig. \ref{fig:RvH-J110}a is lower than the normalized saturation magnetization, reflecting the intermediate magnetization state expected for this field orientation. On the other hand, figures \ref{fig:RvH-J110}b, \ref{fig:RvH-J110}e and \ref{fig:RvH-J110}c, \ref{fig:RvH-J110}f, and their respective insets show the $\rho_l(H)$ and $\rho_t(H)$ measured at 100 K. The insets show the $\rho_l(H)$ and $\rho_t(H)$ in an extended range of -150 mT to 150 mT. The hysteresis and resistivity loops indicate that the sample is saturated at 150 mT. The labels (i) to (v) indicate the regions of the hysteresis and resistance loops to discuss the magnetization process. 

We start our analysis discussing $\rho_{l}$($H$) and $\rho_{t}$($H$) when the field is applied along the hard axis. In the saturation state the magnetization is parallel to the external magnetic field and along the cubic hard axis. As we decrease the magnetic field the magnetization will tend to rotate towards the easy axis [1$\overline{1}$0]. This can be seen in the Kerr loop of Fig. \ref{fig:RvH-J110}a, in which we highlighted 5 different regions of the hysteresis loop: (i) saturation state, (ii) close to the remanence value, (iii) low negative field values, (iv) after magnetization reversal and (v) saturation for the negative field values. We can also identify these 5 regions in the $\rho_l(H)$ and $\rho_t(H)$ plots shown in Figs. \ref{fig:RvH-J110}b and \ref{fig:RvH-J110}c, respectively. From the longitudinal resistance curve we observe that the saturation field  is larger than the one extracted from the hysteresis loop (Inset of Fig. \ref{fig:RvH-J110}a). As the longitudinal resistance provides information about the magnetization component parallel to the current direction, it should not change once the sample is saturated at magnetic field values around $\mu_0H = 150$ mT. Upon decreasing the external field towards zero, the magnetization gradually rotates towards the easy axis [1$\overline{1}$0]. As a result, $\rho_l$ increases because the component of the magnetization parallel to the current direction also increases as the system approaches the remanent state. The magnetization reversal (region iii) occurs at a coercive field of $\mu_0H_\mathrm{coer} \sim 2$ mT, as evidenced by the abrupt decrease in $\rho_l$ at this field, followed by an increase in the signal, which indicates that the magnetization is now close to the [$\overline{1}$10] direction (region iv). For negative magnetic fields beyond the coercive field, the magnetization aligns along the [$\overline{1}$00] direction (region v), resulting in a subsequent decrease in the longitudinal resistance. The same reasoning can be used to describe the behavior of the longitudinal resistance when we increase the external magnetic field from -150 mT to 150 mT. In Fig. \ref{fig:RvH-J110}c, the transverse resistivity $\rho_t$ also provides information on the magnetization process. It is important to notice that $\rho_t (H = 0)$ is not zero in most AMR experiments, commonly due to spurious signals arising from several factors such as contact misalignment or inhomogeneities in the lithography. In our case, we normalized the transverse voltage signal at room temperature by subtracting the offset of the transverse voltage $\rho_t (H = 0)$ at zero field, but still obtained a small transverse resistivity of $\rho_t$ = 97 n$\Omega\cdot$cm when we decrease the temperature down to 100 K. Considering this, we still can also analyze the magnetization relaxation process from the $\rho_t$ plots. At saturation $\phi_M = \phi_H = 45^\circ$, meaning that the magnetization is aligned with the hard axis of \FeCo{65}{35}, and therefore the components parallel and perpendicular to the current direction are equal in magnitude. As the external magnetic field decreases to zero, the component of the magnetization parallel to the current increases, while the perpendicular component decreases. Consequently, at zero field, $\rho_t$ is expected to approach zero (or 97 n$\Omega\cdot$cm in this case) since $\phi_H \sim 0$. When the external magnetic field is further decreased to negative values, the component of the magnetization perpendicular to the current becomes negative until the magnetization suddenly rotates toward the [$\overline{1}$00] direction at the coercive field (regions iii and iv). Finally, the transverse resistivity recovers its initial value because both components of the magnetization are again oriented 45$^\circ$ away from the current direction. In contrast to $\rho_l$, correcting the offset in $\rho_t$ is relevant, since its sign can change, allowing us to easily trace the path followed by the magnetization vector during the magnetization process. The discussion regarding the behavior of the longitudinal and transverse resistances can also be extended to the case where the magnetic field is applied along an easy axis. Figures \ref{fig:RvH-J110}e and \ref{fig:RvH-J110}f display $\rho_l(H)$ and $\rho_t(H)$ measured at $T = 120$ K with the current applied along the cubic and uniaxial easy axis [1$\overline{1}$0]. The $\rho_l(H)$ and $\rho_t(H)$ curves shown in Figs. \ref{fig:RvH-J110}e and \ref{fig:RvH-J110}f are consistent with a magnetization reversal process occurring when the external magnetic field is applied along the cubic–uniaxial easy directions. When the external field is applied along [1$\overline{1}$0] ($\phi_H = 0^\circ$) or its equivalent direction [$\overline{1}$10] ($\phi_H = 180^\circ$), $\rho_l$ remains nearly constant and independent of the magnetic field, since [1$\overline{1}$0] corresponds to the easy axis of the \FeCo{65}{35} layer and the magnetization stays aligned along this direction even at $H = 0$. The two downward peaks in $\rho_l(H)$ correspond to the abrupt magnetization switching at the coercive field. As discussed in the previous section, our samples exhibit a competition between cubic and uniaxial anisotropies, as evidenced by the Kerr loops shown in Figs. \ref{fig:Kerr}a and \ref{fig:Kerr}b. The $\rho_l(H)$ and $\rho_t(H)$ curves obtained with the external magnetic field applied along the cubic easy axis and the uniaxial hard axis are provided in the Supplementary Data file. It is worth noting that the data presented in Figs. \ref{fig:RvH-J110}b and \ref{fig:RvH-J110}c were acquired at 100 K, whereas the data in Figs. \ref{fig:RvH-J110}e and \ref{fig:RvH-J110}f were measured at 120 K. The dependence of $\rho_l$ and $\rho_t$ on the external magnetic field, with the current applied along [1$\overline{1}$0] (easy axis) and the field applied either along [100] (hard axis) or [1$\overline{1}$0] (easy axis), remains nearly unchanged as the temperature varies from 80 K to 300 K. This invariance suggests that both the cubic and uniaxial anisotropy constants are essentially temperature-independent  in our samples within this range. The temperature dependence of $\rho_l(H)$ and $\rho_t(H)$ is provided in the Supplementary Data file. 

\subsection{Model}
\label{sec:Model}

The magnetization process and the magnetotransport properties can be described using a phenomenological model based on the Stoner–Wohlfarth formalism that considers a macroscopic single-domain approximation. This approach assumes that the magnetization rotates coherently for high magnetic field values and that the formation of magnetic domains below the coercive field can not be neglected. To propose an expression of the free energy density that describes the \FeCo{65}{35} magnetic behavior we consider the shape anisotropy, an in-plane uniaxial easy axis and a cubic magnetocrystalline energy. We express this energy density using spherical coordinates ($\theta$, $\phi$) for the representation of the magnetization \textbf{M} and external magnetic field \textbf{H} vectors as shown in Fig. \ref{fig:scheme}. The magnetic free energy of the system can be expressed as:
\begin{equation}
\begin{aligned}
\label{Eq.Energy}
F(\theta_M, \phi_M) = -\mu_0MH \sin \theta_M \cos(\phi_M - \phi_H) + \\ \frac{1}{2}\mu_0 M^2 \cos^2 \theta_M - \frac{K_c}{4} (\sin^2 2\theta_M + \sin^2 2\phi_M \sin^4 \theta_M) \\ - K_u \sin^2\theta_M \cos^2(\phi_M - \phi_u),
\end{aligned}
\end{equation}
where the first term corresponds to the Zeeman energy, the second term to the shape anisotropy, the third term to the uniaxial anisotropy and the last term to the cubic anisotropy. In the expression, $K_c$ and $K_u$ are the anisotropy constants associated to the cubic and uniaxial anisotropy and defined as $K_c = \frac{1}{2} \mu_0M_sH_c$ and  $K_u = \frac{1}{2} \mu_0M_sH_u$, respectively. It is also assumed that the uniaxial axis is in the plane of the sample in a direction given by the angle $\phi_u$. By differentiating with respect to $\theta$ and setting $\frac{\partial F}{\partial \theta} = 0$, we obtain $\theta_M = \frac{\pi}{2}$. After replacing this value of $\theta_M$ in Eq. \ref{Eq.Energy} we obtain a simplified free energy expression in terms of $\phi_M$:
\begin{equation}
\label{eq:F}
F(\phi_M) = -\mu_0MH \cos(\phi_M - \phi_H) - \frac{K_c}{4} \sin^2 2\phi_M - K_u \cos^2(\phi_M - \phi_u).
\end{equation}

By minimizing the free energy given in Eq. \ref{eq:F}, we determined the in-plane equilibrium magnetization angles $\phi_M$ for each value of the external magnetic field $H$ and for different field orientations $\phi_H$. This procedure allows us to obtain the magnetization component either along the external magnetic field—used to simulate the Kerr loops—or along the current direction—used to fit the magnetotransport measurements. To fit the experimental data, the anisotropy constants $K_c$ and $K_u$ were treated as fitting parameters for the normalized Kerr hysteresis loops and the normalized longitudinal and transverse voltages. The Kerr loops obtained from our samples, as well as from our previous studies on different \FeCo{100-x}{x} alloys \cite{Velazquez2020, Velazquez2024, Saba2025}, confirm that the uniaxial anisotropy is weaker than the cubic anisotropy. However, it must still be considered to accurately determine the equilibrium magnetization angles from the total energy expression. In our previous work \cite{Saba2025}, we already discussed the role of the uniaxial anisotropy in fitting the longitudinal and transverse voltage as a function of the external field. The uniaxial term not only breaks the symmetry imposed by the cubic anisotropy—thereby defining the magnetization relaxation path—but also gives rise to a two-step magnetization reversal when the magnetic field is applied along the [110] direction, as shown in Fig. \ref{fig:Kerr}a. Figs. \ref{fig:Kerr}a and \ref{fig:Kerr}b also show the simulated Kerr loops when the external magnetic field is applied along the [110] and [1$\overline{1}$0], respectively. To reproduce the Kerr loops, we employed the anisotropy constants and the magnetization equilibrium angles obtained from the fitting of Eq.~\ref{eq:F}: $\mu_0H_c = (-2.95 \pm 0.06)$ mT, corresponding to $K_c = -2.36$ kJ/m$^3$, and $\mu_0H_u = (2.73 \pm 0.06)$ mT, corresponding to $K_u = 2.18$ kJ/m$^3$, assuming a saturation magnetization of $M_s = 1600$ kA/m. 

\begin{figure}[ht]
\centering
\includegraphics[width=\linewidth]{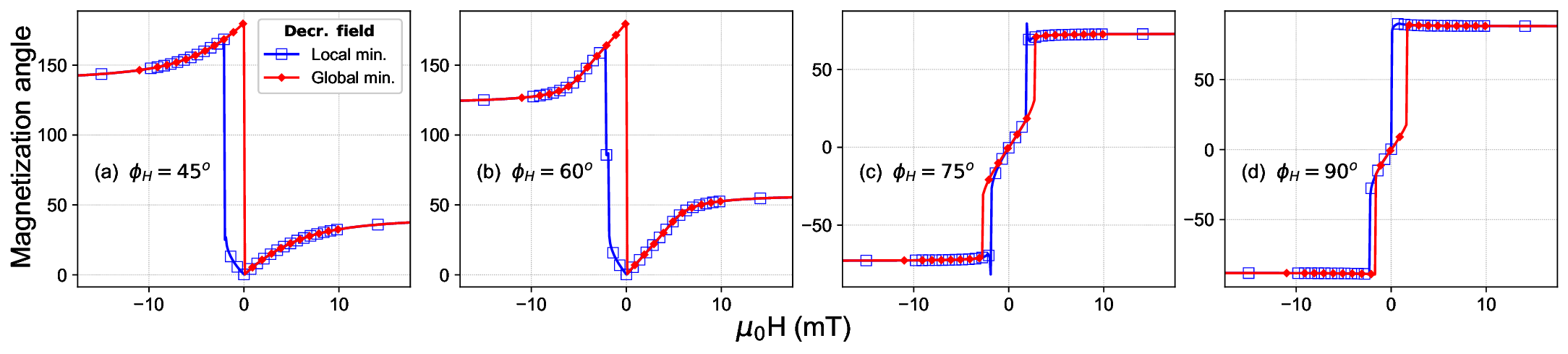}
\caption{Magnetization angle ($\phi_M$) as a function of the magnetic field applied along different directions ($\phi_H$) when the external magnetic field is decreasing from saturation to negative values. Blue squares and red diamonds correspond to the value of $\phi_M$ when the local and the global minima are considered, respectively.}
\label{fig:phiMphiH}
\end{figure}

The fitted parameters accurately reproduce the hysteresis behavior and, more importantly, the resistivity loops measured as a function of the external magnetic field orientation $\phi_H$. For an accurate fitting of the longitudinal and transverse voltage signals, as well as the Kerr loops, both local and global energy minima obtained from the minimization of Eq. \ref{eq:F} were considered. By accounting for these metastable states and assuming a gradual evolution of the magnetization direction, the model successfully reproduces the remanent magnetization and coercive field values observed in both the resistivity and Kerr measurements. Figure~\ref{fig:phiMphiH} shows the equilibrium magnetization angle $\phi_M$, obtained from the minimization of Eq.~\ref{eq:F}, as a function of the external magnetic field applied along different directions ($\phi_H$). 

Figure~\ref{fig:phiMphiH}a ($\phi_H = 45^\circ$) describes the magnetization relaxation process observed in the Kerr loop measured with the external magnetic field applied along the cubic hard axis [100], shown in Fig.~\ref{fig:RvH-J110}a. The calculated $\phi_M$ dependence with the external magnetic field for $\phi_H$ = 90$^\circ$ shown in figure \ref{fig:phiMphiH}d also captures the characteristic two-step magnetization reversal occurring when the field is applied along the [110] direction (Fig. \ref{fig:Kerr}a). The model also captures the gradual evolution of the magnetization for intermediate orientations, such as $\phi_H = 60^\circ$ (Fig.~\ref{fig:phiMphiH}b) and $\phi_H = 75^\circ$ (Fig.~\ref{fig:phiMphiH}c). In the latter case, the equilibrium magnetization angle shows that the reversal proceeds through an abrupt rotation toward $\phi_H$ values near zero under positive field conditions, eventually stabilizing at $\phi_H = 0^\circ$ as the field approaches zero. Since the model does not account for magnetic domain formation, the remanent magnetization for orientations close to $\phi_H = 90^\circ$ cannot be fully reproduced. Overall, these results provide a coherent picture of the magnetization dynamics across different field orientations, underscoring the importance of including metastable states and continuous rotation processes to accurately describe the hysteresis behavior and the magnetotransport response discussed below.

The main sources of uncertainty in the fitting arise from slight misalignment of the lithographically defined Hall bars with respect to the crystallographic axes and between the Hall bars and the applied magnetic field. These effects were considered during the fitting procedure. To account for a possible deviation of the uniaxial anisotropy direction, we introduced the angle $\phi_u$ to describe its orientation. The best fit yielded a small misalignment of $\phi_u \approx -0.6^\circ$.

\subsubsection{Angular dependence of $\rho_l(H)$ and $\rho_t(H)$}
\begin{figure}[ht]
\centering
\includegraphics[width=\linewidth]{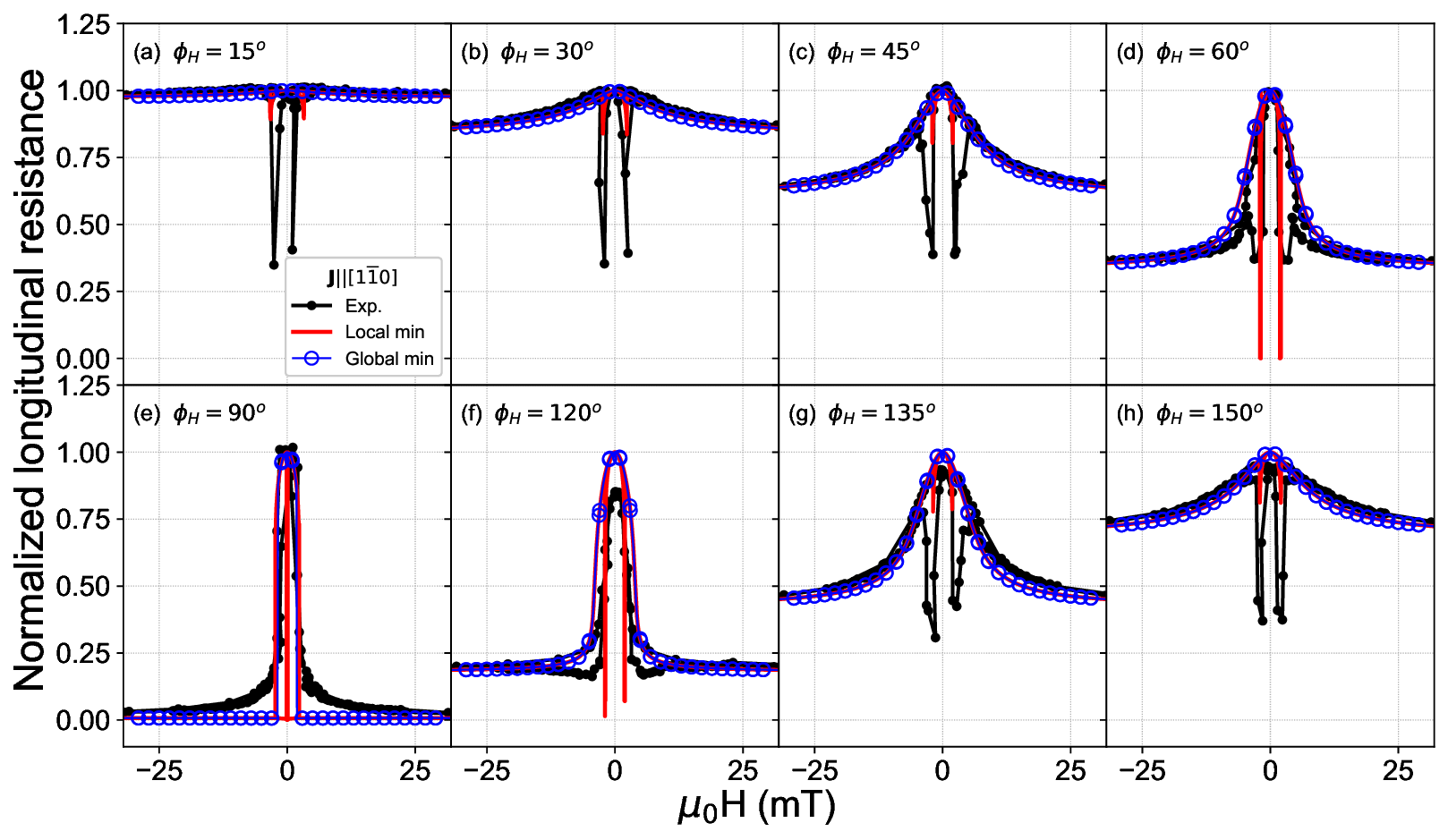}
\caption{Normalized longitudinal voltages as a function of the magnetic field applied along different directions ($\phi_H$). The black circles and line correspond to the experimental data, the red solid line represents the simulated longitudinal resistance calculated using the local minima, and the blue circles correspond to the resistance calculated using the global minima. Measurements were performed with the current applied along the easy axis [1$\overline{1}$0] and were acquired at $T = 80$~K.}
\label{fig:phiH}
\end{figure}

As discussed in the previous section, the longitudinal and transverse voltage measurements were used to determine the magnetization reversal process under external magnetic fields applied along specific in-plane directions ($\phi_H$). Figure \ref{fig:phiH} presents the normalized longitudinal resistance as a function of the external magnetic field for various $\phi_H$ values between 15$^\circ$ and 150$^\circ$, with the current applied along the easy axis [1$\overline{1}$0]. The $\rho_{l}$ values were calculated using Eq. \ref{Eq:rhoxx}, normalized, and plotted as a function of the external magnetic field to enable direct comparison with the simulated longitudinal resistance, obtained from the magnetization equilibrium angles by considering both local and global energy minima. The Supplementary Data file additionally includes the corresponding normalized transverse resistance curves and their respective simulations for the same $\phi_H$ values.

Figures~\ref{fig:phiH}a–\ref{fig:phiH}h show the evolution of the longitudinal resistance as the direction of the external magnetic field is rotated from [1$\overline{1}$0] to [$\overline{1}$10]. As discussed in Figure~\ref{fig:RvH-J110}, when the field is applied along [1$\overline{1}$0] ($\phi_H = 0^\circ$), $\rho_{l}$ remains nearly constant and independent of the magnetic field, since [1$\overline{1}$0] corresponds to the easy axis of \FeCo{65}{35} and the magnetization remains aligned along this direction even at $H = 0$. As $\phi_H$ deviates from the easy-axis direction (e. g. for $\phi_H = 15^\circ$ or $\phi_H = 30^\circ$), the system approaches saturation asymptotically, and the maximum of the normalized longitudinal resistance appears only at $H = 0$, indicating the reorientation of the magnetization toward the easy axis. For angles close to $\phi_H = 45^\circ$, the downward peaks in the longitudinal resistance curves are not abrupt, indicating that the magnetization reorientation is not instantaneous but occurs progressively as the magnetization follows the rotation of the external magnetic field. This gradual transition suggests a smooth evolution between local energy minima rather than an abrupt switching process. As expected, the plots for supplementary $\phi_H$ values—such as $\phi_H = 45^\circ$ and $135^\circ$, $60^\circ$ and $120^\circ$, or $30^\circ$ and $150^\circ$—exhibit the same behavior. The series of plots for different $\phi_H$ values highlights the strong dependence of the magnetotransport response on the magnetic field orientation. By comparing the experimental curves with the simulated longitudinal resistance obtained from the local (red lines) and global (blue circles) minima in Fig.~\ref{fig:phiH}, the magnetization process can be consistently described in the same way as for $\phi_H = 0^\circ$, as discussed in the previous section. As expected, complete magnetic saturation is achieved only at relatively low fields when the external field is applied along the anisotropy axes. For intermediate angles, or for directions slightly deviated from these axes, the magnetization—and consequently the longitudinal resistance—exhibit an asymptotic behavior, reaching saturation only at magnetic field values above approximately 60 mT in some cases. The extended-field-range plots of both the longitudinal and transverse resistances are provided in Section S6 of the supplementary data file.

In Figures~\ref{fig:phiH}a–h, the model accurately reproduces the behavior of the normalized longitudinal resistance for magnetic field values above the saturation field, including the curvature observed near saturation. However, it is important to note that for magnetic field values between –20 mT $< H <$ 20 mT, i.e., near the coercive field, the experimental data deviate significantly from the model. This deviation is expected, as the formation of magnetic domains is not accounted for within the Stoner–Wohlfarth formalism. Despite this limitation, the magnetization process—and, more importantly, the trajectories followed by the magnetization—are well captured across all the measurements. Although domain formation is evidenced by the downward peaks of varying widths, the influence of the local and global energy minima on the magnetization process remains clearly observable. The downward peaks appearing in the longitudinal resistance curves, which correspond to magnetization reversal at the coercive field, can be attributed to the magnetization following a metastable (local) minimum in the energy landscape rather than the global minimum.

\section{Conclusions}
In summary, we investigated the anisotropic magnetoresistance in epitaxial \FeCo{65}{35} thin films, focusing on the influence of crystal symmetry, magnetic field direction, and temperature. The combination of cubic and uniaxial anisotropies determines the magnetization reversal process and leads to a strongly orientation-dependent magnetotransport response. The AMR ratio exhibits clear crystalline anisotropy, reaching 0.155\% and 0.104\% when the current is applied along the hard and easy axes, respectively, representing a 49\% variation. Moreover, the AMR ratio measured along [100] remains nearly temperature-independent, while that along [1$\overline{1}$0] increases by approximately 30\% as the temperature decreases to 80~K, evidencing the influence of spin–orbit coupling and tetragonal distortions on the AMR magnitude. By fitting the Kerr and magnetotransport measurements using a phenomenological model based on the Stoner–Wohlfarth formalism, we accurately reproduced the experimental hysteresis loops and determined the equilibrium magnetization paths. The best agreement with experiment was obtained with cubic and uniaxial anisotropy constants of $K_c = -2.36$ kJ/m$^3$ and $K_u = 2.18$ kJ/m$^3$, respectively. These results demonstrate that epitaxial \FeCo{65}{35} films constitute a promising platform for engineering magnetoresistive devices with tunable anisotropic responses. The AMR magnitude and sensitivity can be optimized by controlling the crystal orientation and operating temperature, paving the way for thermally stable and directionally selective magnetic sensors based on FeCo alloys.

\section*{Acknowledgments}
Technical support from Rubén E. Benavides, César Pérez and Matías Guillén is greatly acknowledged. We thank the Vicerrectorado de Investigacion of the Universidad Nacional de Ingeniería (Peru) and the Formative Research Projects FC-PFR-01-2024 (SPINUNI) and FC-PFR-05-2025 (SPINUNI-II). This work was partially supported by Conicet under Grant PIBAA 2022-2023 (project MAGNETS) Grant ID. 28720210100099CO; ANPCyT Grant PICT 2021-00113 (project DISCO) and U.N. Cuyo Grant 06/C556 from Argentina. L. Avilés-Félix thanks Claudio Ferrari from the Departamento de Micro y Nanotecnología (CNEA - INN) for the preparation of the lithography masks and S. Anguiano for his assistance with the etching.

\bibliographystyle{unsrt}
\bibliography{Biblio}

@PREAMBLE{
 "\providecommand{\noopsort}[1]{}" 
 # "\providecommand{\singleletter}[1]{#1}%" 
}

@article{Thomson1856,
author = {Thomson, William },
title = {XIX. On the electro-dynamic qualities of metals: Effects of magnetization on the electric conductivity of nickel and of iron},
journal = {Proceedings of the Royal Society of London},
volume = {8},
number = {},
pages = {546-550},
year = {1857},
doi = {10.1098/rspl.1856.0144},
URL = {https://royalsocietypublishing.org/doi/abs/10.1098/rspl.1856.0144}}

@article{1970Campbell,
doi = {10.1088/0022-3719/3/1S/310},
url = {https://dx.doi.org/10.1088/0022-3719/3/1S/310},
year = {1970},
publisher = {},
volume = {3},
number = {1S},
pages = {S95},
author = {I A Campbell and A Fert and O Jaoul},
title = {{The spontaneous resistivity anisotropy in Ni-based alloys}},
journal = {Journal of Physics C: Solid State Physics}
}

@article{Freitas1990,
title = {{Anisotropic magnetoresistance in Co films}},
journal = {Journal of Magnetism and Magnetic Materials},
volume = {83},
number = {1},
pages = {113-115},
year = {1990},
issn = {0304-8853},
doi = {https://doi.org/10.1016/0304-8853(90)90452-V},
url = {https://www.sciencedirect.com/science/article/pii/030488539090452V},
author = {P.P. Freitas and A.A. Gomes and T.R. McGuire and T.S. Plaskett}
}

@article{1960Juretschke,
    author = {Juretschke, H. J.},
    title = "{Electromagnetic Theory of dc Effects in Ferromagnetic Resonance}",
    journal = {Journal of Applied Physics},
    volume = {31},
    number = {8},
    pages = {1401-1406},
    year = {1960},
    issn = {0021-8979},
    doi = {10.1063/1.1735851}}

@Article{Berger1988,
  author   = {Berger, L. and Freitas, P. P. and Warner, J. D. and Schmidt, J. E.},
  journal  = {Journal of Applied Physics},
  title    = {On the temperature dependence of the magnetoresistance of ferromagnetic alloys},
  year     = {1988},
  issn     = {0021-8979},
  number   = {10},
  pages    = {5459-5461},
  volume   = {64},
  doi      = {10.1063/1.342347},
  url      = {https://doi.org/10.1063/1.342347},
}

@article{Tondra1993,
    author = {Tondra, Mark and Lottis, Daniel K. and Riggs, K. T. and Chen, Youjun and Dahlberg, E. Dan and Prinz, G. A.},
    title = "{Thickness dependence of the anisotropic magnetoresistance in epitaxial iron films}",
    journal = {Journal of Applied Physics},
    volume = {73},
    number = {10},
    pages = {6393-6395},
    year = {1993},
    issn = {0021-8979},
    doi = {10.1063/1.352607},
    url = {https://doi.org/10.1063/1.352607}}

@article{Mankovsky2010,
  title = {{First-principles calculation of the Gilbert damping parameter via the linear response formalism with application to magnetic transition metals and alloys}},
  author = {Mankovsky, S. and K\"odderitzsch, D. and Woltersdorf, G. and Ebert, H.},
  journal = {Phys. Rev. B},
  volume = {87},
  issue = {1},
  pages = {014430},
  numpages = {11},
  year = {2013},
  publisher = {American Physical Society},
  doi = {10.1103/PhysRevB.87.014430},
  url = {https://link.aps.org/doi/10.1103/PhysRevB.87.014430}}

@article{Kokado2012,
author = {Kokado ,Satoshi and Tsunoda ,Masakiyo and Harigaya ,Kikuo and Sakuma ,Akimasa},
title = {{Anisotropic Magnetoresistance Effects in Fe, Co, Ni, Fe$_4$N, and Half-Metallic Ferromagnet: A Systematic Analysis}},
journal = {Journal of the Physical Society of Japan},
volume = {81},
number = {2},
pages = {024705},
year = {2012},
doi = {10.1143/JPSJ.81.024705}}

@article{Ganguly2014,
    author = {Ganguly, A. and Kondou, K. and Sukegawa, H. and Mitani, S. and Kasai, S. and Niimi, Y. and Otani, Y. and Barman, A.},
    title = {{Thickness dependence of spin torque ferromagnetic resonance in Co$_{75}$Fe$_{25}$/Pt bilayer films}},
    journal = {Applied Physics Letters},
    volume = {104},
    number = {7},
    pages = {072405},
    year = {2014},
    issn = {0003-6951},
    doi = {10.1063/1.4865425},
    url = {https://doi.org/10.1063/1.4865425}}

@Article{Haidar2015,
  author   = {Haidar, S. M. and Iguchi, R. and Yagmur, A. and Lustikova, J. and Shiomi, Y. and Saitoh, E.},
  journal  = {Journal of Applied Physics},
  title    = {{Reducing galvanomagnetic effects in spin pumping measurement with Co$_{75}$Fe$_{25}$ as a spin injector}},
  year     = {2015},
  issn     = {0021-8979},
  number   = {18},
  pages    = {183906},
  volume   = {117},
  doi      = {10.1063/1.4921359},
  url      = {https://doi.org/10.1063/1.4921359}}

@Article{Schoen2016,
  author   = {Schoen, Martin A. W. and Thonig, Danny and Schneider, Michael L. and Silva, T. J. and Nembach, Hans T. and Eriksson, Olle and Karis, Olof and Shaw, Justin M.},
  title    = {Ultra-low magnetic damping of a metallic ferromagnet},
  doi      = {10.1038/nphys3770},
  issn     = {1745-2481},
  number   = {9},
  pages    = {839--842},
  url      = {https://doi.org/10.1038/nphys3770},
  volume   = {12},
  journal  = {Nature Physics},
  refid    = {Schoen2016},
  year     = {2016}}

@article{Harder2016,
title = {Electrical detection of magnetization dynamics via spin rectification effects},
journal = {Physics Reports},
volume = {661},
pages = {1-59},
year = {2016},
issn = {0370-1573},
doi = {https://doi.org/10.1016/j.physrep.2016.10.002},
url = {https://www.sciencedirect.com/science/article/pii/S0370157316303167},
author = {Michael Harder and Yongsheng Gui and Can-Ming Hu}}

@article{Li2019,
  title = {{Giant Anisotropy of Gilbert Damping in Epitaxial CoFe Films}},
  author = {Li, Yi and Zeng, Fanlong and Zhang, Steven S.-L. and Shin, Hyeondeok and Saglam, Hilal and Karakas, Vedat and Ozatay, Ozhan and Pearson, John E. and Heinonen, Olle G. and Wu, Yizheng and Hoffmann, Axel and Zhang, Wei},
  journal = {Phys. Rev. Lett.},
  volume = {122},
  issue = {11},
  pages = {117203},
  numpages = {6},
  year = {2019},
  publisher = {American Physical Society},
  doi = {10.1103/PhysRevLett.122.117203},
  url = {https://link.aps.org/doi/10.1103/PhysRevLett.122.117203}}

@article{Weber2019,
doi = {10.1088/1361-6463/ab2096},
url = {https://dx.doi.org/10.1088/1361-6463/ab2096},
year = {2019},
publisher = {IOP Publishing},
volume = {52},
number = {32},
pages = {325001},
author = {Ramon Weber and Dong-Soo Han and Isabella Boventer and Samridh Jaiswal and Romain Lebrun and Gerhard Jakob and Mathias Kläui},
title = {{Gilbert damping of CoFe-alloys}},
journal = {Journal of Physics D: Applied Physics}}

@article{Zeng2020,
  title = {{Intrinsic Mechanism for Anisotropic Magnetoresistance and Experimental Confirmation in ${\mathrm{Co}}_{x}{\mathrm{Fe}}_{1-x}$ Single-Crystal Films}},
  author = {Zeng, F. L. and Ren, Z. Y. and Li, Y. and Zeng, J. Y. and Jia, M. W. and Miao, J. and Hoffmann, A. and Zhang, W. and Wu, Y. Z. and Yuan, Z.},
  journal = {Phys. Rev. Lett.},
  volume = {125},
  issue = {9},
  pages = {097201},
  numpages = {7},
  year = {2020},
  publisher = {American Physical Society},
  doi = {10.1103/PhysRevLett.125.097201},
  url = {https://link.aps.org/doi/10.1103/PhysRevLett.125.097201}}

@article{Zeng2020a,
doi     = {10.1088/1367-2630/abb16d},
url     = {https://dx.doi.org/10.1088/1367-2630/abb16d},
year    = {2020},
publisher = {IOP Publishing},
volume  = {22},
number  = {9},
pages   = {093047},
author  = {Fanlong Zeng and Xi Shen and Yi Li and Zhe Yuan and Wei Zhang and Yizheng Wu},
title   = {{Role of crystalline and damping anisotropy to the angular dependences of spin rectification effect in single crystal CoFe film}},
journal = {New Journal of Physics}}

@article{Velazquez2020,
title = {{Relaxation mechanisms in ultra-low damping Fe$_{80}$Co$_{20}$ thin films}},
author = {D. Vel\'azquez-Rodriguez and J. E. G\'omez and G. Alejandro and L. Avil\'es-F\'elix and M. van Landeghem and E. Goovaerts and A. Butera},
journal = {Journal of Magnetism and Magnetic Materials},
volume = {504},
pages = {166692},
year = {2020},
issn = {0304-8853},
doi = {https://doi.org/10.1016/j.jmmm.2020.166692}}

@Article{Velazquez2021,
author    = {Vel\'azquez Rodriguez, D. and G\'omez, J. E. and Morbidel, L. and Costanzo Caso, P. A. and Milano, J. and Butera, A.},
title     = {{High spin pumping efficiency in Fe$_{80}$Co$_{20}$/Ta bilayers}},
doi       = {10.1088/1361-6463/ac02fc},
issn      = {0022-3727},
number    = {32},
pages     = {325002},
url       = {https://dx.doi.org/10.1088/1361-6463/ac02fc},
volume    = {54},
journal   = {Journal of Physics D: Applied Physics},
publisher = {IOP Publishing},
year      = {2021}}

@article{Natale2021,
  title = {Field-dependent nonelectronic contributions to thermal conductivity in a metallic ferromagnet with low Gilbert damping},
  author = {Natale, M. R. and Wesenberg, D. J. and Edwards, Eric R. J. and Nembach, Hans T. and Shaw, Justin M. and Zink, B. L.},
  journal = {Phys. Rev. Mater.},
  volume = {5},
  issue = {11},
  pages = {L111401},
  numpages = {7},
  year = {2021},
  month = {Nov},
  publisher = {American Physical Society},
  doi = {10.1103/PhysRevMaterials.5.L111401},
  url = {https://link.aps.org/doi/10.1103/PhysRevMaterials.5.L111401}
}

@article{Lim2022,
    author = {Lim, Byeonghwa and Mahfoud, Mohamed and Das, Proloy T. and Jeon, Taehyeong and Jeon, Changyeop and Kim, Mijin and Nguyen, Trung-Kien and Tran, Quang-Hung and Terki, Ferial and Kim, CheolGi},
    title = {Advances and key technologies in magnetoresistive sensors with high thermal stabilities and low field detectivities},
    journal = {APL Materials},
    volume = {10},
    number = {5},
    pages = {051108},
    year = {2022},
    issn = {2166-532X},
    doi = {10.1063/5.0087311}}

@article{Velazquez2024,
doi = {10.1088/1361-6463/ad5b6d},
year = {2024},
publisher = {IOP Publishing},
volume = {57},
number = {39},
pages = {395003},
author={Vel\'azquez Rodriguez, Daniel and G\'omez, Javier E and Avil\'es-F\'elix, Luis and Ampuero Torres, Jose Luis and Torres, Teobaldo E. and P\'erez Mart\'inez, \'Angel and Morbidel, Leonardo and Goijman, Dafne Yael and Rojas-Sanchez, Juan-Carlos and Aguirre, Myriam and Milano, Juli\'an and Butera, Alejandro},
title={Intrinsic and extrinsic relaxation mechanisms for controlling spin current intensity in {Fe$_{100-x}$Co$_x$/Ta} bilayers},
journal = {Journal of Physics D: Applied Physics}}

@Article{Wang2024,
author ="Wang, Xiaotao and Guo, Lin and Bezsmertna, Olha and Wu, Yuhan and Makarov, Denys and Xu, Rui",
title  ="Printed magnetoresistive sensors for recyclable magnetoelectronics",
journal  ="J. Mater. Chem. A",
year  ="2024",
volume  ="12",
issue  ="37",
pages  ="24906-24915",
publisher  ="The Royal Society of Chemistry"}

@article{Tamulynas2026,
title = {{Temperature stability and compensation of AMR sensors in practical applications}},
journal = {AEU - International Journal of Electronics and Communications},
volume = {203},
pages = {156082},
year = {2026},
issn = {1434-8411},
author = {Matas Tamulynas and Eidenis Kasperavičius and Vytautas Markevičius and Dangirutis Navikas and Mindaugas Žilys and Algimantas Valinevičius and Michal Frivaldsky and Roman Sotner and Jan Jerabek and Darius Andriukaitis},}

@article{Dong2023,
  title = {Anisotropic magnetoresistance due to magnetization-dependent spin-orbit interactions},
  author = {Dong, M. Q. and Guo, Zhi-Xin and Wang, X. R.},
  journal = {Phys. Rev. B},
  volume = {108},
  issue = {2},
  pages = {L020401},
  numpages = {6},
  year = {2023},
  publisher = {American Physical Society},
  doi = {10.1103/PhysRevB.108.L020401}}

@article{Zhang2025,
doi = {10.1088/1402-4896/adce5f},
url = {https://dx.doi.org/10.1088/1402-4896/adce5f},
year = {2025},
publisher = {IOP Publishing},
volume = {100},
number = {5},
pages = {055976},
author = {Zhang, Mingsong and Peng, Bin and Yuan, Zhe and Zhang, Wenxu},
title = {{Fourfold anisotropic magnetoresistance in FeCo and Fe$_4$N}},
journal = {Physica Scripta}}

@article{Miao2024,
  title = {Anisotropic Galvanomagnetic Effects in Single Cubic Crystals: A Theory and Its Verification},
  author = {Miao, Yu and Sun, Junwen and Gao, Cunxu and Xue, Desheng and Wang, X. R.},
  journal = {Phys. Rev. Lett.},
  volume = {132},
  issue = {20},
  pages = {206701},
  numpages = {7},
  year = {2024},
  publisher = {American Physical Society},
  doi = {10.1103/PhysRevLett.132.206701},
  url = {https://link.aps.org/doi/10.1103/PhysRevLett.132.206701}}

@article{Saba2025,
doi = {10.1088/1361-6463/adf11b},
url = {https://dx.doi.org/10.1088/1361-6463/adf11b},
year = {2025},
publisher = {IOP Publishing},
volume = {58},
number = {30},
pages = {305005},
author = {Saba, L and Gómez, J E and Pérez-Morelo, D J and Anguiano, S and Velázquez Rodriguez, D and Butera, A and Granada, M and Avilés-Félix, L},
title = {{Magnetization process in epitaxial Fe$_{85}$Co$_{15}$ thin films via anisotropic magnetoresistance}},
journal = {Journal of Physics D: Applied Physics}}

@article{Natale2024,
  title = {{Magnon-drag and field-direction dependent thermopower in low-damping ferromagnetic ${\mathrm{Co}}_{25}{\mathrm{Fe}}_{75}$ alloy thin films}},
  author = {Natale, M. R. and Wesenberg, D. J. and Zink, B. L.},
  journal = {Phys. Rev. Mater.},
  volume = {8},
  issue = {4},
  pages = {044402},
  numpages = {11},
  year = {2024},
  publisher = {American Physical Society},
  doi = {10.1103/PhysRevMaterials.8.044402},
  url = {https://link.aps.org/doi/10.1103/PhysRevMaterials.8.044402}
}

@article{Solis2025,
	author={Solís León, Raúl and Garcia Manuz, Ines and Chaluvadi, Sandeep Kumar and Pierron, Victor and Polewczyk, Vincent and Petrov, Yu and Vinai, Giovanni and Pautrat, Alain and Ajejas, Fernando and Guillet, Bruno and Flament, Stephane and Orgiani, Pasquale and Mechin, Laurence and Perna, Paolo},
	title={{Low-noise anisotropic magnetoresistance sensing at human body temperature: Unveiling the optimal doping in La$_\mathrm{1-x}$Sr$_\mathrm{x}$MnO$_\mathrm{3}$ films}},
	journal={Journal of Physics D: Applied Physics},
	url={http://iopscience.iop.org/article/10.1088/1361-6463/ae18de},
	year={2025}}

@Article{Miklusis2022,
AUTHOR = {Miklusis, Donatas and Markevicius, Vytautas and Navikas, Dangirutis and Ambraziunas, Mantas and Cepenas, Mindaugas and Valinevicius, Algimantas and Zilys, Mindaugas and Okarma, Krzysztof and Cuinas, Inigo and Andriukaitis, Darius},
TITLE = {{Erroneous Vehicle Velocity Estimation Correction Using Anisotropic Magnetoresistive (AMR) Sensors}},
JOURNAL = {Sensors},
VOLUME = {22},
YEAR = {2022},
NUMBER = {21},
ARTICLE-NUMBER = {8269},
URL = {https://www.mdpi.com/1424-8220/22/21/8269},
PubMedID = {36365966},
ISSN = {1424-8220},
DOI = {10.3390/s22218269}
}

@ARTICLE{Balamutas2023,
  author={Balamutas, Juozas and Navikas, Dangirutis and Markevičius, Vytautas and Čepėnas, Mindaugas and Valinevičius, Algimantas and Žilys, Mindaugas and Frivaldsky, Michal and Li, Zhixiong and Andriukaitis, Darius},
  journal={IEEE Access}, 
  title={Passing Vehicle Road Occupancy Detection Using the Magnetic Sensor Array}, 
  year={2023},
  volume={11},
  number={},
  pages={50984-50993},
  doi={10.1109/ACCESS.2023.3278986}}

@ARTICLE{Ripka2013,
	author = {Ripka, Pavel and Butta, Mattia and Platil, Antonin},
	title = {Temperature stability of AMR sensors},
	year = {2013},
	journal = {Sensor Letters},
	volume = {11},
	number = {1},
	pages = {74 – 77},
	doi = {10.1166/sl.2013.2807},
	url = {https://www.scopus.com/inward/record.uri?eid=2-s2.0-84878001176&doi=10.1166%2fsl.2013.2807&partnerID=40&md5=7ed86a8791d4814c094d613217fc4cf7},
	type = {Article},
	publication_stage = {Final}}

@article{NICOLICEA2025,
title = {Flexible anisotropic magnetoresistive sensors for novel eddy current testing applications},
journal = {Measurement},
volume = {253},
pages = {117340},
year = {2025},
issn = {0263-2241},
doi = {https://doi.org/10.1016/j.measurement.2025.117340},
url = {https://www.sciencedirect.com/science/article/pii/S0263224125006992},
author = {Alberto Nicolicea and Eduardo Sergio Oliveros-Mata and Denys Makarov and Michael Melzer and Matthias Pelkner}}

@article{LIN2025,
title = {Design and implementation of a nondestructive testing system for magnetic field imaging based on machine learning},
journal = {Journal of Magnetism and Magnetic Materials},
volume = {614},
pages = {172638},
year = {2025},
issn = {0304-8853},
doi = {https://doi.org/10.1016/j.jmmm.2024.172638},
url = {https://www.sciencedirect.com/science/article/pii/S0304885324009296},
author = {Ming-Yi Lin and Ou-Wen Lee}
}

@Article{Cheng2025,
AUTHOR = {Cheng, Jen-Chieh and You, Min-Chang and Anbalagan, Aswin kumar and Su, Guang-Yang and Chuang, Kai-Wei and Yang, Chao-Yao and Lee, Chih-Hao},
TITLE = {Investigation of a Magnetic Sensor Based on the Magnetic Hysteresis Loop and Anisotropic Magnetoresistance of CoFe Thin Films Epitaxial Grown on Flexible Mica and Rigid MgO Substrates with Strain Effect},
JOURNAL = {Micromachines},
VOLUME = {16},
YEAR = {2025},
NUMBER = {4},
ARTICLE-NUMBER = {412},
URL = {https://www.mdpi.com/2072-666X/16/4/412},
PubMedID = {40283288},
ISSN = {2072-666X},
DOI = {10.3390/mi16040412}
}

@article{ZENG2020c,
title = {{Strong current-direction dependence of anisotropic magnetoresistance in single crystalline Fe/GaAs(110) films}},
journal = {Journal of Magnetism and Magnetic Materials},
volume = {499},
pages = {166204},
year = {2020},
issn = {0304-8853},
doi = {https://doi.org/10.1016/j.jmmm.2019.166204},
url = {https://www.sciencedirect.com/science/article/pii/S0304885319330057},
author = {F.L. Zeng and C. Zhou and M.W. Jia and D. Shi and Y. Huo and W. Zhang and Y.Z. Wu}}

@article{Zhang2023,
doi = {10.1088/1361-648X/ace01c},
url = {https://doi.org/10.1088/1361-648X/ace01c},
year = {2023},
publisher = {IOP Publishing},
volume = {35},
number = {39},
pages = {395803},
author = {Zhang, Mingsong and Peng, Bin and Zhang, Wanli and Zhang, Wenxu},
title = {{Effect of atomic anti-site disorder on the anisotropic magnetoresistance in Fe$_{50}$Co$_{50}$ alloys}},
journal = {Journal of Physics: Condensed Matter}
}

@article{LEE2024,
title = {{Composition dependence of the orbital torque in Co$_x$Fe$_{1-x}$ and Ni$_x$Fe$_{1-x}$ alloys: Spin-orbit correlation analysis}},
journal = {Current Applied Physics},
volume = {67},
pages = {60-68},
year = {2024},
issn = {1567-1739},
doi = {https://doi.org/10.1016/j.cap.2024.07.014},
url = {https://www.sciencedirect.com/science/article/pii/S1567173924001718},
author = {Hojun Lee and Hyun-Woo Lee}
}

@Article{Bo2025,
  author   = {Bo, Guohao and Xu, Xiaoguang and Meng, Kangkang and Wu, Yong and Jiang, Yong},
  title    = {Flexible magnetic films and spintronic devices},
  doi      = {10.1038/s44306-025-00113-z},
  issn     = {2948-2119},
  number   = {1},
  pages    = {48},
  url      = {https://doi.org/10.1038/s44306-025-00113-z},
  volume   = {3},
  journal  = {npj Spintronics},
  refid    = {Bo2025},
  year     = {2025},
}
\end{document}